\documentclass[aps,pra,twocolumn,english,superscriptaddress,floatfix,longbibliography]{revtex4-2}

\usepackage[english]{babel}
\usepackage{textcomp,xcolor,natbib}
\usepackage{dcolumn}
\usepackage{graphicx}
\graphicspath{{Figures/}}

\usepackage{amsmath, bbm}
\usepackage{latexsym}
\usepackage{amsfonts}   
\usepackage{amssymb}
\usepackage{array}      
\usepackage{epsfig}
\usepackage{txfonts}
\usepackage{xcolor,braket}
\usepackage[colorlinks=true,linkcolor=blue,urlcolor=blue,citecolor=blue]{hyperref}
\usepackage[normalem]{ulem}
\usepackage{soul}
\usepackage{dsfont}

\usepackage{xfrac}  
\usepackage{relsize}

\usepackage{amsthm}
\usepackage{mathrsfs, mathtools, amsmath}

\usepackage[normalem]{ulem}
\usepackage{cancel}
\usepackage{enumitem}

\usepackage{xspace, comment}

\newcommand{\abs}[1]{\lvert #1 \rvert}

\renewcommand{\Re}{\operatorname{Re}}
\renewcommand{\Im}{\operatorname{Im}}

\newcommand{\omegarrc}{\omega_{\text{rrc}}}

\definecolor{lavender}{rgb}{0.71, 0.49, 0.86}
\definecolor{olivgreen}{rgb}{0.73, 0.72, 0.42}

\usepackage{enumitem}

\newcommand{\bignone}{}
\newcommand{\mathd}{\mathrm{d}}
\newcommand{\mathe}{\mathrm{e}}

\newcommand{\tmop}[1]{\ensuremath{\operatorname{#1}}}

\begin{document}

\title{Synchronization effects in a periodically driven two-level system}

\author{Federico Settimo}
\email{fesett@utu.fi}
\affiliation{Department of Physics and Astronomy,
University of Turku, FI-20014 Turun yliopisto, Finland}

\author{Bassano Vacchini}
\email{bassano.vacchini@mi.infn.it}
\affiliation{Dipartimento di Fisica ``Aldo Pontremoli'', Universit{\`a} degli Studi di Milano, Via Celoria 16, I-20133 Milan, Italy}
\affiliation{Istituto Nazionale di Fisica Nucleare, Sezione di Milano, Via Celoria 16, I-20133 Milan, Italy}

\begin{abstract}
We study phase–synchronization in a driven two–level system coupled to a non-Markovian bosonic reservoir.
The dynamics is described by treating the system–bath coupling and the coherent drive without invoking the rotating–wave approximation, and simulated using the numerically exact hierarchical equations of motion.
We observe that a robust phase–locking develops and that the corresponding synchronization measure rapidly acquires a finite value when the system is tuned to what we identify as a resonant-ratio condition, namely when the ratio between the drive amplitude and its frequency coincides with a zero of the Bessel function $J_0$. We provide an explanation for this phenomenon by means of a static approximation derived from a Fourier analysis of the periodically driven Hamiltonian.
\end{abstract}


\maketitle
\section{Introduction}

The role of external interventions in the dynamics of classical and quantum systems has been the subject of intense investigation, not only because such interventions are often unavoidable, but also due to the variety of new effects they may induce.
While there is no conceptual obstruction to treating classical systems as perfectly isolated, this is no longer true for quantum systems, for which any degree of observation or control unavoidably modifies their dynamics. In this respect, the treatment of open quantum systems plays a central role, see e.g. \cite{Breuer2002,Rivas2012,Vacchini2024} and references therein.
External interventions may play a dual role: on the one hand they may act as disturbances that need to be suppressed or corrected, while on the other hand they may induce effects that are qualitatively new and desired. This is particularly relevant when the interaction combines coherent and incoherent components.
An important class of coherent interventions is provided by periodic driving, with the interplay between such driving and environmental disturbances that still remains to be fully understood, and has been the focus of several recent investigations \cite{Zhao2010a,Colla2021a,NafariQaleh2022a,Cao2023a,Dann2025a}.

Here we consider in particular how the interplay between a non-Markovian environment and a coherent periodic drive determines phase-locking and synchronization effects in a two-level system.
The very concept of synchronization in these systems was initially criticized \cite{Roulet2018a}, although its physical relevance was later recognized \cite{Parra-Lopez2020a,Xiao2023a}, and it has also been experimentally demonstrated \cite{Krithika2022a,Zhang2023a}.
These studies are part of the broader effort of investigating synchronization phenomena in quantum systems, arising either from entrainment with respect to an external signal \cite{Walter2014a,Chen2025a,Almani2025a}, or from self-organization \cite{Koppenhofer2019a,Lorenzo2022a,Zhang2020a,Galve2017a,Giorgi2016a,Benedetti2016a,Giorgi2013a,Karpat2019a,Karpat2020a,Manzano2013a,Schmolke2022a}.
Since the inception of the notion, a large body of work has been devoted to synchronization in quantum two-level systems, based either on weak-coupling approximations of the environmental noise effects \cite{Chen2025a} or on exactly solvable models where a rotating-wave approximation is adopted \cite{Almani2025a}. In particular, the impact of non-Markovian dynamics in the onset of synchronization has been investigated \cite{Zhou2021a,Ali2022a,Ali2024a,Huang2023a,Karpat2021a,Karpat2025a}.
In this work, we consider the exact dynamics of a two-level system coupled to a bosonic bath, without invoking a rotating-wave approximation either in the system–bath interaction or in the driving term, so that non-Markovian effects can be fully taken into account. This is feasible by employing a numerically exact technique, known as hierarchical equations of motion (HEOM) \cite{Tanimura2020a,Johansson2012a,Johansson2013a}, which provides an efficient strategy for computing the reduced dynamics of the open system also in the presence of driving.

The manuscript is organized as follows.
In Sect.~\ref{sect:mic} we introduce the model, specifying the features of the coherent driving and of the incoherent environmental disturbance. For the coherent part we exploit the periodicity of the driving and perform a suitable change of frame that makes explicit the different operator-valued Fourier components. For the incoherent contribution we consider a non-Markovian bosonic bath, allowing for an exact treatment without further approximations thanks to the HEOM technique.
In Sect.~\ref{sec:results} we introduce the figures of merit commonly used in the literature to assess synchronization effects and present the results obtained by studying their dependence on the relevant model parameters. In particular, we analyze the emergence of limit cycles in a phase-space representation of the system state and quantify the impact of the ratio between driving strength and driving frequency in triggering a positive synchronization measure.
Finally, in Sect.~\ref{sect:end} we summarize our findings and outline possible future developments.

\section{Model of non-Markovian driven system}\label{sect:mic}

In this Section, we introduce the model of a periodically driven two-level
system interacting with a bosonic environment, which serves as the
paradigmatic setting for investigating dissipative quantum dynamics beyond the
Markovian regime. The combined effect of coherent driving and environmental
coupling gives rise to a rich interplay between coherent control and memory
effects, that we will investigate in order to explore the appearance of
quantum synchronization phenomena.

\subsection{Two-level driven system and high frequency approximation}

We consider a two-level quantum system subject to a periodic driving field,
described by the Hamiltonian
\begin{eqnarray}
  H (t) & = & \frac{\hbar \omega_0}{2} \sigma_z + \frac{\Delta}{2} \sigma_x +
  \frac{\hbar \Omega}2 \cos (\omega t) \sigma_x .  \label{eq:ht}
\end{eqnarray}
Here, $\hbar \omega_0$ denotes the energy splitting between the ground and
excited states, $\hbar \Omega$ the driving strength, $\omega$ the driving
frequency determining the periodicity, and $\Delta$ a bias term that can be
interpreted as a static driving. This time-dependent Hamiltonian captures the
essential ingredients of coherent control by means of a classical field and
appears in the description of a variety of physical systems
{\cite{Grifoni1998a,Magazzu2018a}}. Despite its apparent simplicity, the
system does not allow for a closed analytical solution.

It can, however, be conveniently analyzed by applying a time-dependent unitary
transformation corresponding to a change of reference frame. To this aim, we
note that the Hamiltonian can be written as
\begin{eqnarray}
  H (t) & = & V +f (\omega t)  W
\end{eqnarray}
with $f$ a periodic function, allowing for a systematic expansion in powers of
the inverse driving frequency
{\cite{Pegg1973a,Goldman2014a,TwyeffortIrish2022a}}. We consider the unitary
transformation to a rotating frame
\begin{eqnarray}
  U_r (t) & = & \mathe^{- \frac{i}{\hbar} W \int_0^t \mathd \tau \bignone f
  (\omega \tau)}, 
\end{eqnarray}
so that the system wavefunction in the new frame, $\psi_r$, obeys the
Schr{\"o}dinger equation
\begin{eqnarray}
  i \hbar \frac{\mathd}{\tmop{dt}} \psi_r (t) & = & H_r (t) \psi_r (t) 
\end{eqnarray}
with
\begin{eqnarray}
  H_r (t) & = & U_r (t)^{\dag} H (t) U_r (t) - i \hbar U_r (t)^{\dag}
  \frac{\mathd}{\tmop{dt}} U_r (t) . 
\end{eqnarray}
The resulting Hamiltonian $H_r (t)$ remains periodic with period $T = 2 \pi /
\omega$, and can thus be expanded in operator-valued Fourier components as
\begin{eqnarray}
  H_r (t) & = & \sum_{n = - \infty}^{+ \infty} \bignone \mathcal{H}_n
  \mathe^{- in \omega t}, 
\end{eqnarray}
where
\begin{eqnarray}
  \mathcal{H}_n & = & \frac{1}{T} \int_0^T \mathd t \, \bignone U_r (t)^{\dag}
  VU_r (t) \mathe^{+ in \omega t}.
\end{eqnarray}
The considered transformation thus allows one to write the Hamiltonian as a sum of contributions that take into account the effect of the periodic drive on the dynamics.

For the explicit expression in Eq.~(\ref{eq:ht}) we obtain
\begin{eqnarray}
  U_r (t) & = & \mathe^{-i \frac{\Omega}{2 \omega} \sin (\omega t) \sigma_x}, 
\end{eqnarray}
and, using the identity
\begin{eqnarray}
  \mathe^{i \frac{\alpha}{2} \sigma_x } \sigma_z \mathe^{- i \frac{\alpha}{2}
  \sigma_x} & = & \cos (\alpha) \sigma_z + \sin (\alpha) \sigma_y 
\end{eqnarray}
together with the Bessel integral representation
\begin{eqnarray}
  J_n (x) & = & \frac{1}{2 \pi} \int_0^{2 \pi} \mathd \theta \,\mathe^{- ix \sin
  (\theta)} \, \bignone \mathe^{+ in \theta}, 
\end{eqnarray}
where $J_n$ denotes the Bessel function of the first kind of integer order $n$
{\cite{Gradshteyn1965}}, which satisfies
\begin{eqnarray}
  J_{- n} (x) & = & (- 1)^n J_n (x), 
\end{eqnarray}
we thus obtain the following expressions for the even and odd operator-valued
Fourier coefficients
\begin{gather}\label{eq:H_mod_2k}
  \mathcal{H}_{2 k}  =  \frac{\hbar\omega_0}{2} J_{2 k} \left(
  \frac{\Omega}{\omega} \right) \sigma_z+\delta_{k,0}\,\frac\Delta2\sigma_x \\ \label{eq:H_mod_2k+1}
  \mathcal{H}_{(2 k + 1)}  = i \frac{\hbar\omega_0}{2} J_{2 k + 1}
  \left( \frac{\Omega}{\omega} \right) \sigma_y
\end{gather}
where $k \in \mathbbm{Z}$. This leads to the following compact expression for the
Hamiltonian in the rotating frame:
\begin{eqnarray}
\label{eq:H_mod_r}
  H_r (t) & = & \frac{\hbar \omega_0}{2} J_0 \left( \frac{\Omega}{\omega}
  \right) \sigma_z + \frac{\Delta}{2} \sigma_x \nonumber\\
  &  & + \hbar \omega_0 \sum_{k = 1}^{+ \infty} \bignone J_{2 k} \left(
  \frac{\Omega}{\omega} \right) \cos [(2 k) \omega t] \sigma_z  \\ \nonumber
  &  & + \hbar \omega_0 \sum_{k =0}^{+ \infty} \bignone J_{2 k + 1} \left(
  \frac{\Omega}{\omega} \right) \sin [(2 k + 1) \omega t] \sigma_y . 
\end{eqnarray}
In this expression we have made explicit the static contribution, which
differs from the original static term in Eq.~(\ref{eq:ht}) only by a
renormalization of the spectral gap by a factor given by the zeroth-order
Bessel function evaluated at the ratio of the driving strength to the driving
frequency. This relevant correction has already been observed in several
physical systems {\cite{Shirley1965a2,Holthaus1992a,Grifoni1998a}} and provides a good approximation to the dynamics whenever the driving
frequency or the driving strength are large compared to the original energy gap $\hbar\omega_0$.

\subsection{Non-Markovian modeling of bosonic bath}

The dissipative environment is modeled as a bosonic reservoir linearly
coupled to the system, corresponding to the well-known spin--boson model
{\cite{Leggett1987a}}. The influence of the environment is fully characterized
by its spectral density, which determines both the strength and the timescales
of the resulting memory effects. To capture the ensuing non-Markovian
dynamics beyond perturbative or Markovian approximations, we employ the HEOM approach {\cite{Tanimura2020a}}, a
numerically exact technique that allows us to resolve the system's dynamics
across different regimes of coupling and temperature. In the present case, the
HEOM method can be conveniently implemented using the QuTiP package
{\cite{Johansson2012a,Johansson2013a}}, which provides an efficient numerical
framework for open quantum systems.

The total system--environment Hamiltonian is taken in the form
\begin{equation}\label{eq:Htot}
    \begin{split}
        H_{SE} (t) = & \frac{\hbar \omega_0}{2} \sigma_z + \frac{\Delta}{2}
  \sigma_x + \frac{\hbar \Omega}2 \cos (\omega t) \sigma_x\\
  &+ \sum_k \hbar \omega_k
  a^{\dagger}_k a_k + \sigma_x  \sum_k g_k  (a_k + a^{\dagger}_k), 
    \end{split}
\end{equation}
where $a^{\dagger}_k$ and $a_k$ denote, respectively, the creation and
annihilation operators of the $k$-th mode of the environmental bosonic degrees
of freedom with frequency $\omega_k$, while $g_k$ represents the coupling
strength of this mode to the system. Note that the rotating-wave approximation
is applied neither to the driving nor to the system--environment coupling
terms. Owing to the linearity of the interaction, the relevant features of the
environmental coupling can be fully captured by means of the spectral density,
formally defined as
\begin{eqnarray}
  J (\omega) & = & \sum_k | g_k |^2 \delta (\omega - \omega_k) \, , 
\end{eqnarray}
which we take in the Drude--Lorentz form,
\begin{eqnarray}
  J (\omega) & = & 2 \lambda \frac{\gamma \omega}{\omega^2 + \gamma^2} \, ,
\end{eqnarray}
where $\gamma$ sets the cut-off frequency and $\lambda$ determines the overall
coupling strength. This standard choice combines an Ohmic (i.e. linear)
behavior at small frequencies with a cut-off at higher frequencies, thus
providing a realistic and numerically convenient description of dissipation.
The specific shape of the cut-off function, while not crucial in determining
the physical behavior, allows for a smooth suppression of high-frequency
contributions. We emphasize that the HEOM approach enables the exact inclusion
of non-Markovian effects in the dynamical evolution, that, in the present case, will mainly depend on the temperature of the environmental state and on the value of the cut-off frequency, offering a comprehensive
description of the driven open quantum system across all relevant timescales.

\section{Results}
\label{sec:results}

We now turn to the analysis of synchronization effects in the dynamics of the
driven dissipative two-level system under consideration. In particular, we aim
at clarifying under which conditions synchronization phenomena are enhanced and
how they manifest at the level of the system's observables. As previously mentioned, we rely on the HEOM technique, which
allows us to access the full non-Markovian dynamics without introducing any
a priori approximation on either the system--bath coupling or the structure of
the driving field. This
numerically exact perspective will be complemented by a comparison between the
observed signatures and the analytical results derived from the microscopic
model presented in Sec.~\ref{sect:mic}.

\begin{figure}
    \centering
    \includegraphics[width=\linewidth]{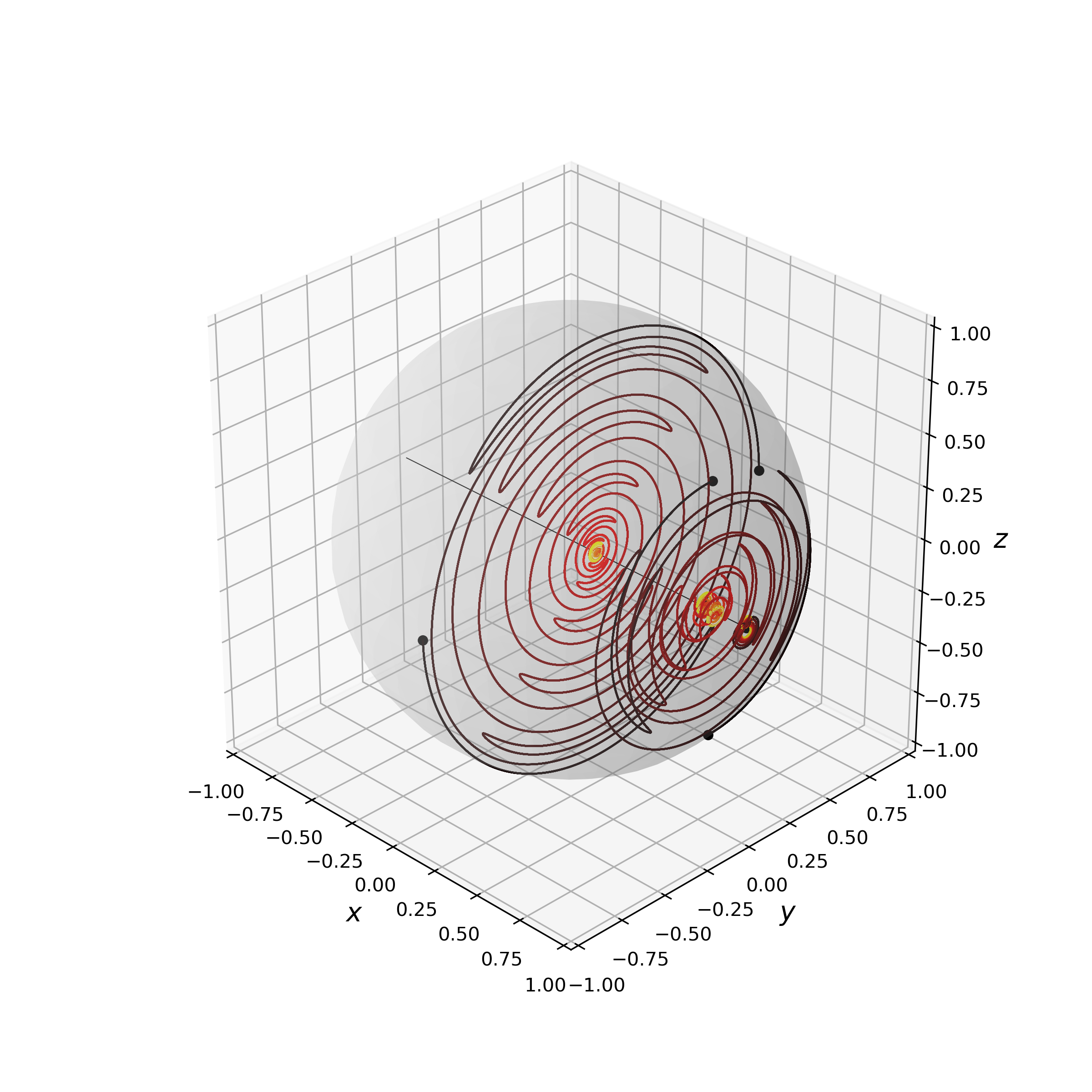}
    \caption{Time evolution of the reduced system under the Hamiltonian of Eq.~\eqref{eq:Htot} for an environment initially in a thermal state, plotted as trajectories on the Bloch sphere.
    The model parameters are set to $\Delta=0$, $\Omega=60\,\omega_0$, $\gamma=\omega_0/2$, $\lambda=\omega_0$, and $T=0.5$.
    The final time of the simulation is $\omega_0\, t_{\rm max}=30$.
    Lighter colors correspond to later times, and the starting point of each trajectory is denoted with a dot.
    The trajectories converge to a limit cycle around the $x$ axis retaining the same $x$ component as the initial state.}
    \label{fig:trajs}
\end{figure}

\subsection{Figures of merit for quantum synchronization}\label{sect:qs}
The concept of synchronization for a two-level quantum system can be naturally
introduced by resorting to a phase--space description associated with the
Hilbert space $\mathbb{C}^2$. In particular, one considers spin--coherent
states $\psi(\theta,\varphi)$ defined as eigenvectors of the spin operator
along the direction $\mathbf{n}$ identified by the polar and azimuthal angles
$\theta$ and $\varphi$, respectively, namely
\begin{eqnarray}
  (\hat{\mathbf{S}}\!\cdot\! \mathbf{n}) \, \psi(\theta,\varphi) = 
  \psi(\theta,\varphi),
\end{eqnarray}
where
\begin{eqnarray}
  \mathbf{n} = (\sin\theta\cos\varphi,\,\sin\theta\sin\varphi,\,\cos\theta), 
  \qquad 
  \hat{\mathbf{S}} = \frac{\hbar}{2}\,\boldsymbol{\sigma},
\end{eqnarray}
with $\boldsymbol{\sigma}=(\sigma_x,\sigma_y,\sigma_z)$ the vector of Pauli matrices.
These states form an overcomplete set satisfying
\begin{eqnarray}
  \frac{1}{2\pi}\!\int_0^{\pi}\!\!\!\mathrm{d}\theta\,\sin\theta
  \int_0^{2\pi}\!\!\!\mathrm{d}\varphi\,
  |\psi(\theta,\varphi)\rangle\langle\psi(\theta,\varphi)|
  = \mathbbm{1},
\end{eqnarray}
thus enabling a phase--space representation of the dynamics on the unit
sphere. For a time-dependent state $\rho(t)$, one introduces the Husimi
$Q$--function
\begin{eqnarray}
\label{eq:Q}
  Q(\theta,\varphi,t) = \frac{1}{2\pi}
  \langle \psi(\theta,\varphi)|\,\rho(t)\,\psi(\theta,\varphi)\rangle,
\end{eqnarray}
which defines a normalized probability density on the $(\theta,\varphi)$
sphere and is such that $0\le Q(\theta,\varphi,t)\le1/2\pi$. The evolution of the state itself can be described through the Bloch
vector $\mathbf{m}(t)$ defined via
\begin{eqnarray}
\label{eq:traiet}
  \rho(t)=\tfrac{1}{2}\bigl(\mathbbm{1}+\mathbf{m}(t)\!\cdot\!\boldsymbol{\sigma}\bigr),
\end{eqnarray}
whose representation on the Bloch sphere directly allows for a visualization of trajectories, as done in Fig.~\ref{fig:trajs}.

\begin{figure}
    \centering
    \includegraphics[width=\linewidth]{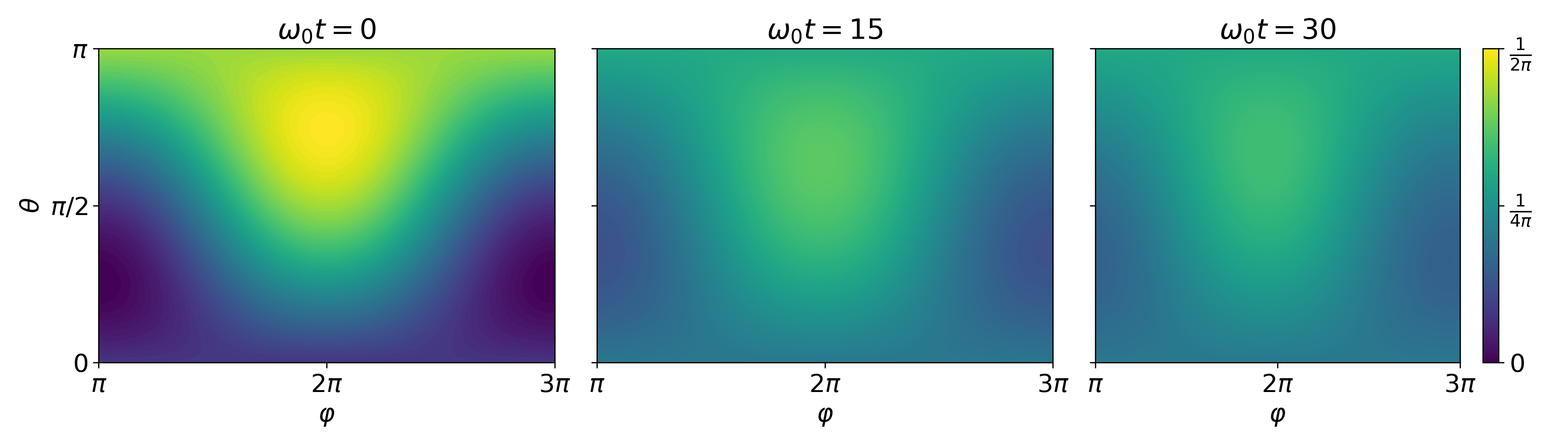}\\
    \includegraphics[width=\linewidth]{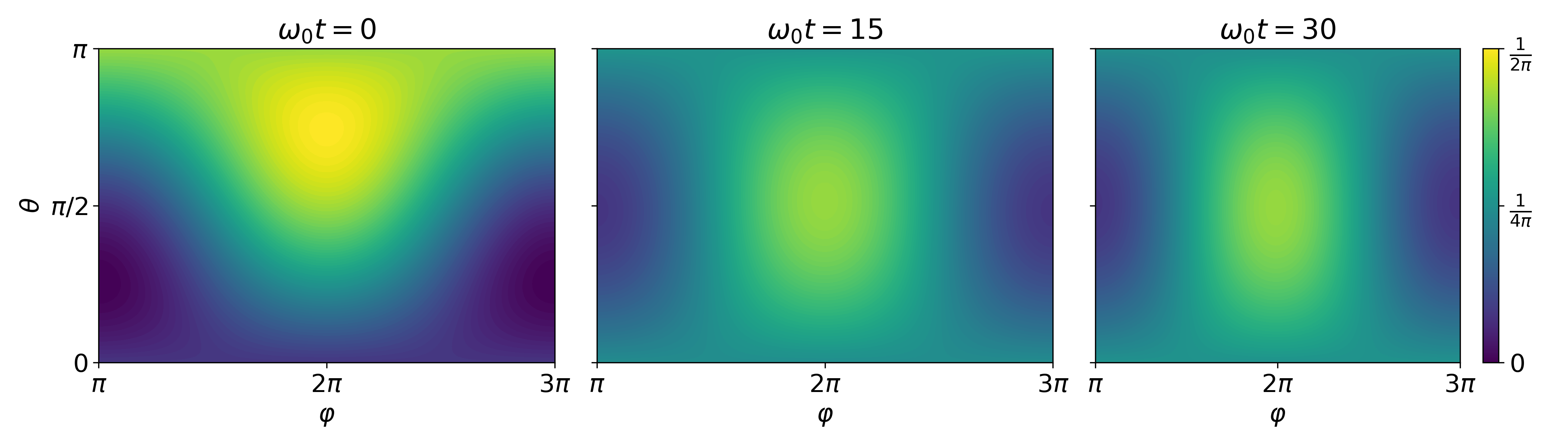}\\
    \includegraphics[width=\linewidth]{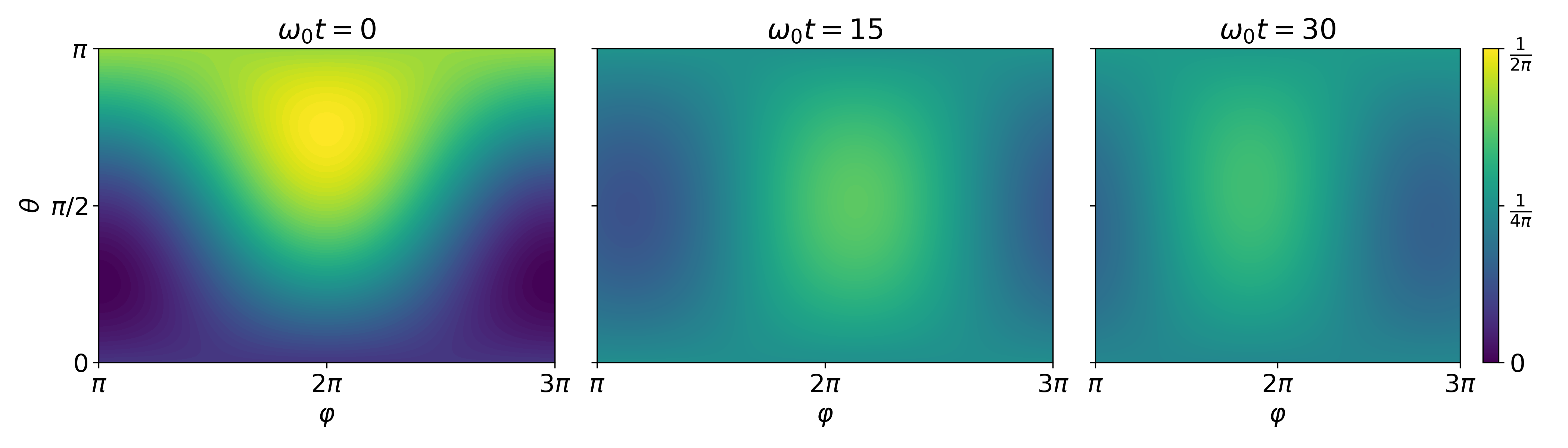}\\
    \caption{Dynamics of $Q$ as a function of the angles $\theta$ and $\varphi$. Top to bottom: $\omega = \omegarrc-\delta\omega$, $\omega=\omegarrc$, and $\omega = \omegarrc+\delta\omega$.
    Left to right: initial, intermediate, and final time. We fix $\delta\omega = 10\omega_0$. The other parameters are the same as in Fig.~\ref{fig:trajs}.}
    \label{fig:Q}
\end{figure}

\begin{figure}
    \centering
    \includegraphics[width=\linewidth]{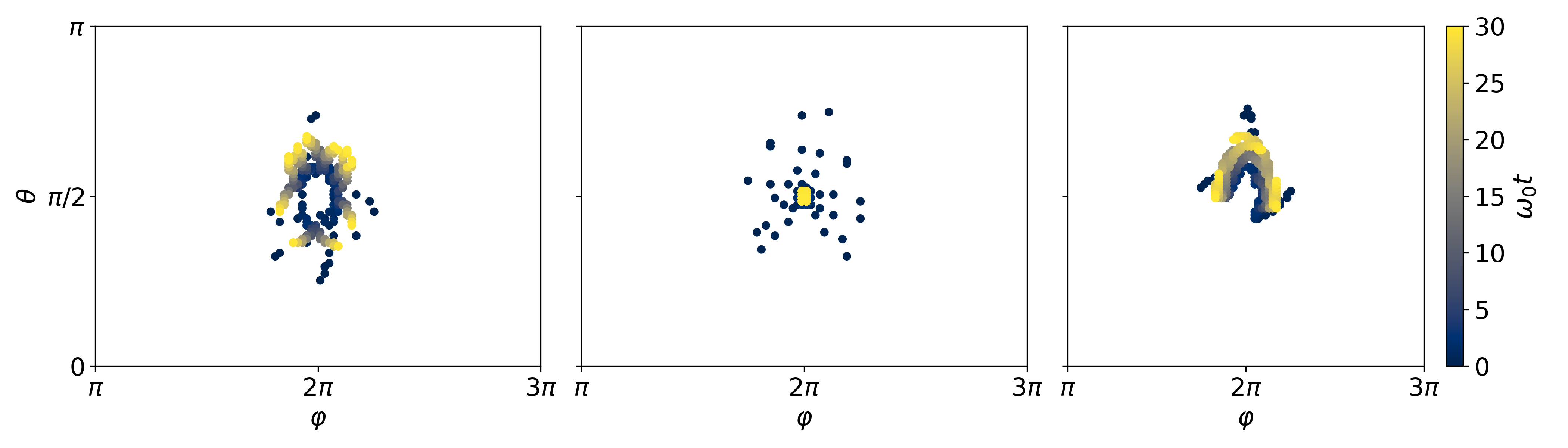}
    \caption{Time evolution of the angles $\theta$, $\varphi$ maximizing $Q$. Time is encoded in the color (gray) gradient of the points, with darker tones corresponding to earlier times. Left to right: $\omega=\omegarrc-\delta\omega$, $\omega=\omegarrc$, $\omega=\omegarrc+\delta\omega$.}
    \label{fig:Q_max}
\end{figure}

Starting from the Husimi
$Q$--function, and in analogy with classical notions of
limit--cycle phase-locking, the following synchronization measure was
introduced in Refs.~\cite{Roulet2018a,Parra-Lopez2020a}:
\begin{eqnarray}
\label{eq:S}
  S(\varphi,t) = \int_{0}^{\pi}\!\mathrm{d}\theta\,\sin\theta\,
  Q(\theta,\varphi,t) - \frac{1}{2\pi},
\end{eqnarray}
obtained by marginalizing over $\theta$ and subtracting the value for a uniform azimuthal
distribution.
We have $\abs{S(\varphi,t)}\le1/8$, with a nonvanishing value signaling the emergence of azimuthal symmetry breaking, i.e. the onset of quantum synchronization.
It is instructive to rewrite this quantity explicitly in terms of the
density matrix elements 
\begin{eqnarray}
  \rho(t)=\begin{pmatrix}
    p(t) & c(t)\\
    \overline{c(t)} & 1-p(t)
  \end{pmatrix},
\end{eqnarray}
where $p(t)$ is the excited--state population and $c(t)$ the coherence. One
then finds the equivalent expression
\begin{eqnarray}
    \label{eq:S_re_im_c}
  S(\varphi,t)=\frac{1}{4}\bigl(
  \Re\{c(t)\}\cos\varphi - \Im\{c(t)\}\sin\varphi
  \bigr),
\end{eqnarray}
which makes explicit the fact that synchronization is rooted in the presence of
persistent coherences.
A nonzero asymptotic value of $S(\varphi,t)$ indicates phase-locking and hence synchronization in the long--time regime.
We recall that alternative approaches to quantum synchronization exist, most
notably those based on entrainment, where one quantifies the locking of a
distinguished system observable to the phase of an external signal~\cite{Cabot2021a}.  The phase--space
criterion adopted here is instead of geometric nature: it detects
synchronization through an azimuthal symmetry breaking in the steady-state
distribution on the Bloch sphere~\cite{Zhang2023a}. The two perspectives are complementary, the
former emphasizing frequency locking at the level of observables, the latter
capturing the structural deformation of the stationary state in phase space.

\begin{figure}
    \centering
    \includegraphics[width=\linewidth]{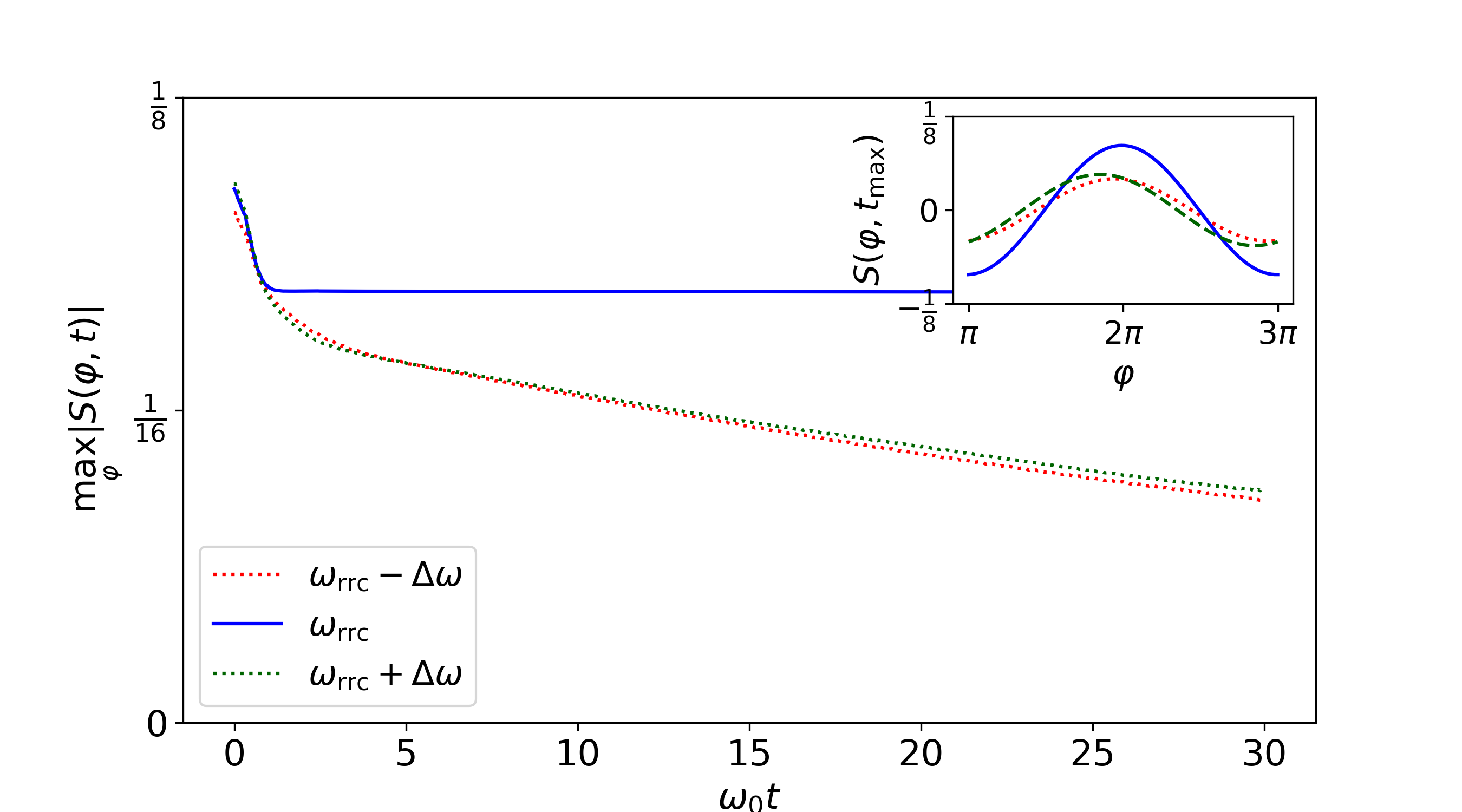}
    \caption{Time evolution of $\max_\varphi \abs{S(\varphi,t)}$ for $\omega=\omegarrc$ and $\omega=\omegarrc\pm\delta\omega$.
    Inset: $S(\varphi,t)$ as a function of $\varphi$ evaluated at the final time $t=t_{\text{max}}$.}
    \label{fig:S_max}
\end{figure}

\subsection{Behavior of trajectories and phase-space quantities}

\begin{figure*}
    \centering
    \includegraphics[width=\linewidth]{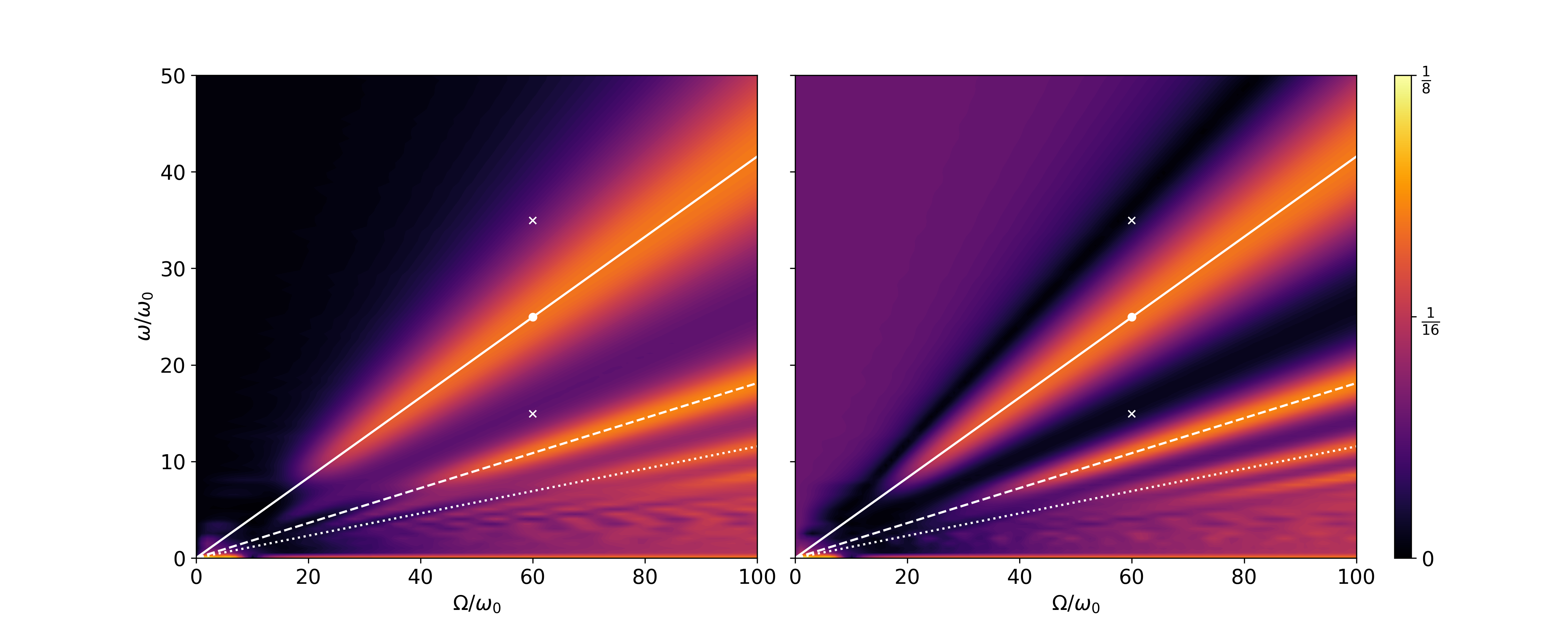}
    \caption{Maximum of $\abs{S(\varphi, t_{\text{max}})}$ as a function of $\Omega$ and $\omega$.
    The white lines are the first three RRCs $\omega = \omegarrc, \omegarrc^2,\omegarrc^3$. The white dots and crosses correspond to the parameters used in Figs.~\ref{fig:Q}-\ref{fig:S_max}.
    In order to improve the stability, the simulation is done by taking an average around $t_{\text{max}}$, in a time window comparable to the duration of a limit cycle.
    Left panel: $\Delta = 0$, right panel: $\Delta = 0.2\hbar\omega_0$.}
    \label{fig:S_Omega_omega}
\end{figure*}

We first analyze the time evolution of the trajectories associated with the
Bloch vector $\mathbf{m}(t)$ of Eq.~\eqref{eq:traiet} on the Bloch sphere. This
provides a direct visualization of the long-time effects induced by the
driving, and allows us to clarify their dependence on the initial condition.
In Fig.~\ref{fig:trajs} we show the dynamics generated by the Hamiltonian of Eq.~\eqref{eq:Htot} for different initial states.
The code used for the simulation is available at \cite{github}.
Motivated by the exact
rotating-frame expression of Eq.~\eqref{eq:H_mod_r}, we investigate how these
trajectories depend on the ratio between the driving strength and the driving
frequency.
In particular, we focus on the static approximation obtained by retaining only
the time-independent term in Eq.~\eqref{eq:H_mod_r}, corresponding to the
Hamiltonian $\mathcal H_0$ in Eq.~\eqref{eq:H_mod_2k}. This approximation is
expected to be accurate in the regime $\omega,\Omega\gg \max\{\omega_0,\Delta\}$,
and predicts the vanishing of the effective Hamiltonian when the ratio
$\Omega/\omega$ coincides with a zero of the Bessel function $J_0$.
We are thus led to introduce a \emph{resonant-ratio condition} (RRC)
\begin{eqnarray}
\label{eq:rrc}
    \omegarrc^k = \frac{\Omega}{z_k},
\end{eqnarray}
where $z_k$ denotes the $k$-th zero of $J_0$. We refer to the first such
resonance $\omegarrc^1$ simply as $\omegarrc$.
In the regime where the static approximation holds, this resonant condition
corresponds to a degeneracy of the Floquet quasienergies. Indeed, due to the
periodicity of the Hamiltonian of Eq.~\eqref{eq:ht}, its dynamics can be analyzed
within Floquet theory, which introduces Floquet eigenvectors and their
associated quasienergies, and provides an explicit representation of the time
evolution operator \cite{Holthaus2015a}.
The trajectories shown in Fig.~\ref{fig:trajs} are obtained under the RRC
condition $\omega=\omegarrc$, starting from different initial states.
All trajectories converge to an asymptotic limit cycle aligned almost perpendicular to the $x$-axis of the Bloch sphere.
The resulting limit cycle clearly breaks the azimuthal symmetry,
hinting at the onset of synchronization induced by the combined action
of the drive and the environment.
Furthermore, the limit cycle has the same $x$ component as the initial state, and therefore the real part of $c(t)$ is preserved, while the imaginary part is suppressed.
According to Eq.~\eqref{eq:S_re_im_c} the limit cycle will have non-zero asymptotic synchronization due to preservation of $\Re\{c(t)\}$.

Motivated by the existence of such limit cycle, we evaluate the figures of merit introduced in Sect.~\ref{sect:qs} to estimate phase-locking and synchronization effects in the resulting dynamics.
In Fig.~\ref{fig:Q}, we present the values of the Husimi-$Q$ function of Eq.~\eqref{eq:Q} at the initial, intermediate, and final time.
The non-uniformity of the $Q$ function with respect to $\theta$ signals the presence of synchronization.
In order to assess the relevance of the RRC, we consider the behavior of $Q$ for three different values of the frequency: at the RRC ($\omega=\omegarrc$) and detuned by an amount $\delta\omega$ ($\omega=\omegarrc\pm\delta\omega$).
For the RRC, the non-uniformity of $Q(\theta,\varphi)$ is larger compared to the other cases, in which it is almost flat.
This non-uniformity of $Q(\theta,\varphi)$ is a strong indication that synchronization is enhanced by the RRC.

Further insight into the dependence of the Husimi–$Q$ function of Eq.~\eqref{eq:Q} on the driving parameters is provided in Fig.~\ref{fig:Q_max}, where we track the values of $\theta$ and $\varphi$ at which the function attains its maximum over time.
In the RRC (middle panel), after a short transient the maximum stabilizes at $\theta = \pi/2$ and $\varphi = 0 \ (\text{mod}\ 2\pi)$, which corresponds to a coherent state lying on the equatorial plane and signals a robust breaking of azimuthal symmetry.
Away from this resonant condition (left and right panels), however, both $\theta$ and $\varphi$ display persistent and large oscillations around their long–time values, even at late times. This behavior reflects the absence of a stable phase–locked regime and confirms that the emergence of a stationary phase reference is tied to the resonant ratio of driving strength and frequency.

It is important to stress that all the quantities used for estimating phase-locking and synchronization have been evaluated in a regime of coupling strength $\lambda$ and cut-off frequency $\gamma$ that correspond to the deep non-Markovian regime for the considered two-level dissipative system~\cite{Clos2012a,Cao2021a}.


\begin{figure}
    \centering
    \includegraphics[width=\linewidth]{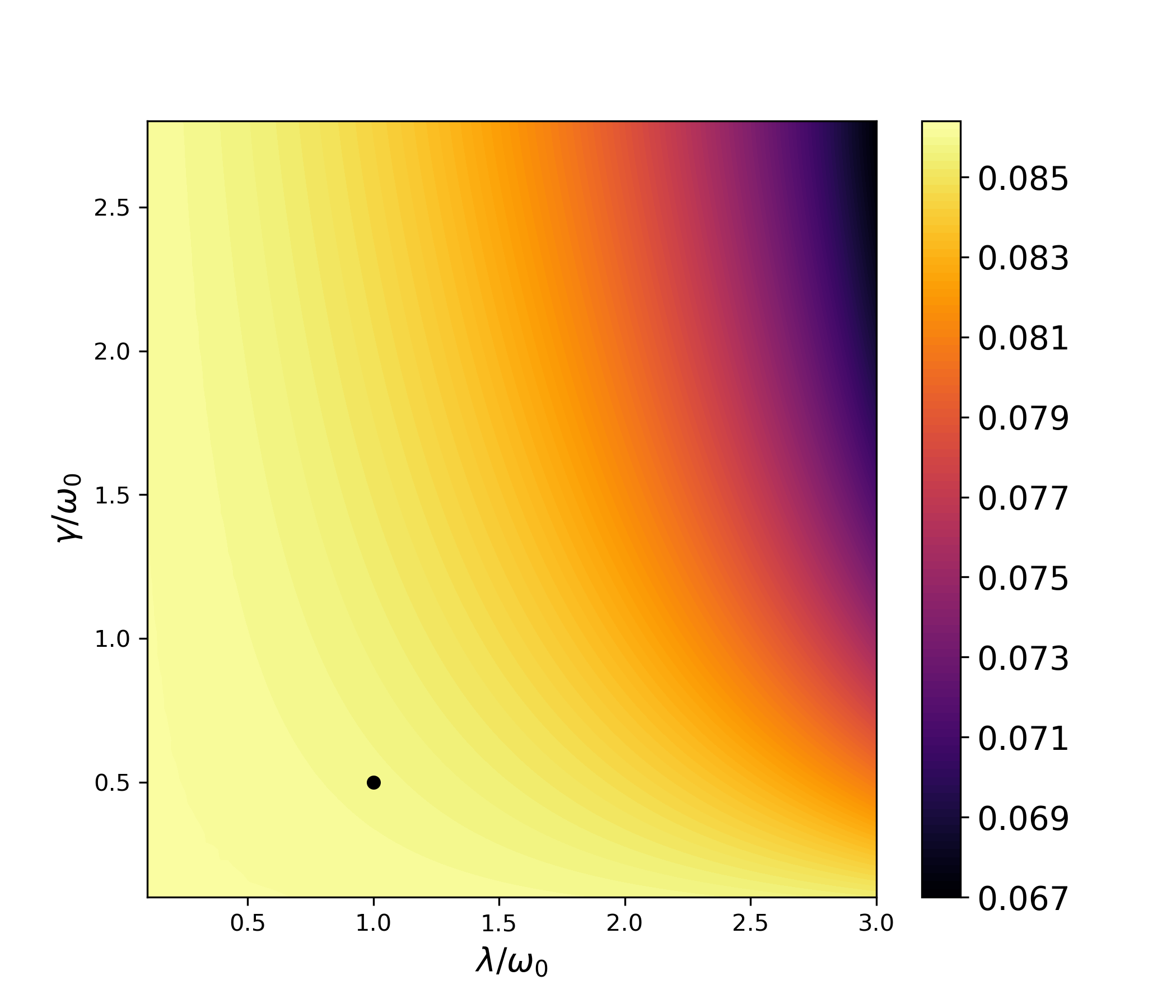}
    \caption{Dependence of the maximum of $S(\varphi, t_{\text{max}})$ in the RRC as a function of the bath coupling $\lambda$ and cut-off frequency $\gamma$.
    The black dot represents the value of the parameters considered in our analysis.}
    \label{fig:bath}
\end{figure}

\subsection{Dependence on model parameters}

We now show that the observed phase-locking and synchronization behavior, although strongly connected to the RRC, is only weakly affected by other bath parameters. To this end, we focus on the 
behavior of the synchronization measure $S(\varphi)$ of Eq.~\eqref{eq:S}, both at intermediate  and
asymptotic times,  while varying the relevant model parameters, though always remaining in the non-Markovian regime.
In Fig.~\ref{fig:S_max} we consider its behavior in time  for the RRC as well as higher and lower driving frequencies. It clearly appears a rapid onset of a non-zero value of the synchronization measure, which remains stable in time for the RRC and rapidly decreases otherwise, hinting again at the role of the RRC in the emergence of synchronization.
This behavior justifies considering a reference time $t_{\text{max}}$ for the study of long-time properties. It is chosen such that in the RRC the system has reached its asymptotic state, while out of resonance the synchronization is almost completely suppressed.
In the inset, we show that the maximum of $S$ for long times is obtained for $\varphi=0$, so that $\cos\varphi=1$ and the weight of the preserved quantity $\Re\{c(t)\}$ in Eq.~\eqref{eq:S_re_im_c} is maximized.

In view of this behavior, we study in detail the behavior of this quantity in the joint ($\Omega,\omega$) dependence.
In the left panel of Fig.~\ref{fig:S_Omega_omega}, we plot the maximum of $S(\varphi,t)$ at the final simulation time $t_{\text{max}}$ as a function of the driving amplitude $\Omega$ and frequency $\omega$. In the region where the static approximation is valid, i.e. for $\Omega,\omega\gg\omega_0$, the synchronization measure exhibits clear maxima along the RRCs $\omega = \omegarrc^k$, indicated by the white lines.
The white dot (crosses) correspond to the parameter sets $\omega=\omegarrc$ ($\omega=\omegarrc\pm\delta\omega$) used in Figs.~\ref{fig:Q}–\ref{fig:S_max}, illustrating that the observed behavior depends solely on the RRC, that is on the ratio $\Omega/\omega$, and not on the absolute values of the two parameters.
Moreover, higher-order resonances $\omegarrc^{k>1}$, shown as dashed and dotted white lines, display a similar enhancement for sufficiently large $\Omega$.
Overall, synchronization is strongly enhanced whenever the RRC Eq.~\eqref{eq:rrc} is satisfied, i.e. when $\Omega/\omega$ coincides with a zero of the Bessel function $J_0$.
We recall that this condition admits a natural interpretation in Floquet theory: it corresponds to the degeneracy of the Floquet quasienergies, a regime which has been shown to underpin a number of distinct physical phenomena, including most recently the characterization of non-Markovian effects~\cite{Follia2025a} and decoherence-protection mechanisms~\cite{Huang2021a,Benhayoune-Khadraoui2025a,Briseno-Colunga2025a,Shadfar2025a}.
In the right panel of Fig.~\ref{fig:S_Omega_omega}, we consider the same figure of merit, but considering a non-zero bias $\Delta$. In the limit of a small bias $\abs\Delta\ll\hbar\omega_0$, it has the effect of changing the validity of the static approximation to the region $\omega,\Omega\gg\sqrt{\omega_0^2+(\Delta/\hbar)^2}$.
In the presence of this non-trivial constant driving $\Delta\sigma_x/2$, a new region with non-zero synchronization emerges for $\omega>\omegarrc$, which however 
presents a much weaker signal compared to the RRC.

The emergence of asymptotic non-vanishing synchronization in the RRC can be interpreted in light of the full system environment Hamiltonian of Eq.~\eqref{eq:Htot}, that upon validity of the static approximation in the rotated reference frame reads
\begin{equation}\label{eq:Htot_static}
    \begin{split}
        H_{SE} (t) = & \frac{\hbar \omega_0}{2} J_0\left(\frac\Omega\omega\right)\sigma_z + \frac{\Delta}{2}
  \sigma_x\\
  &+ \sum_k \hbar \omega_k
  a^{\dagger}_k a_k + \sigma_x  \sum_k g_k  (a_k + a^{\dagger}_k).
    \end{split}
\end{equation}
At the RRC, the first term is zero, and therefore the system Hamiltonian and the coupling with the bath commute.
When this happens, the dynamics is of the dephasing type \cite{Chruscinski2022a}, with the dephasing basis given by the eigenstates of $\sigma_x$.
Therefore, as noted in Fig.~\ref{fig:trajs}, the $x$ component of the Bloch vector, or equivalently $\Re\{c(t)\}$, is preserved, thus explaining the survival of synchronization in the long-time limit.
Off-resonance, instead, the commutativity is broken due to the term proportional to $J_0(\Omega/\omega)\,\sigma_z$, so that the $x$ component of the Bloch vector is no longer preserved, leading to a weaker synchronization.

It is worth stressing that the undriven system environment Hamiltonian, namely Eq.~\eqref{eq:Htot} for $\Omega=0$, gives rise to a non-trivial dynamics which is not simply of dephasing type.
Furthermore, the emergence of dephasing is not solely due to the strength $\Omega$ of the driving, but arises because of a non-trivial interplay between driving strength and frequency, so that it is indeed a resonance feature which gives rise to the synchronization observed in the RRC.
From the point of view of Floquet theory, non-degenerate quasienergies correspond to a $\sigma_z$ driving term, breaking the commutativity with the coupling as well as the preservation of the coherence.
In the RRC the quasienergies become degenerate, thus restoring commutativity and the dynamics is of dephasing type.
Our results do not depend on the particular interaction Hamiltonian, but any coupling of the form $\sigma_x\otimes B$, with $B$ a self-adjoint environmental operator, would give rise to the same effect.

Finally, in Fig.~\ref{fig:bath}, we present the maximum of $S(\varphi, t_{\text{max}})$ as a function of the bath coupling strength $\lambda$ and cut-off frequency $\gamma$, in the RRC and for $\Delta = 0$.
It is possible to notice that the synchronization is a monotonically decreasing function of both parameters, but the dependence is weak, and therefore the default parameters $\gamma = \omega_0/2$, $\lambda =\omega_0$ considered so far do not significantly impact the behavior of the synchronization.
The behavior remains qualitatively the same also for small but non-zero bias $\Delta$.
This behavior can be expected, since in the RRC changing the environmental parameters only alters the timescales over which the dephasing takes place, but not the preservation of $\Re\{c(t)\}$ which gives rise to the synchronization.

\section{Conclusions}\label{sect:end}

We have investigated the emergence of phase synchronization in a periodically driven two-level system coupled to a non-Markovian bosonic environment.
By treating both the system–bath interaction and the coherent drive exactly, i.e. without applying any rotating-wave approximation, we have shown that synchronization arises robustly when the drive frequency $\omega$ and drive strength $\Omega$ satisfy a RRC, defined by the zeros of the Bessel function $J_0$.

This resonant condition marks the point where the effective static Hamiltonian vanishes in the rotating frame, leading to a commutation between the system Hamiltonian and the bath coupling.
Consequently, the resulting dynamics leads to preservation of the system coherences.
The resulting stationary state thus retains a finite degree of coherence and exhibits distinct phase-locking signatures in the long-time limit, clearly showing the emergence of synchronization.

Our analysis highlights the crucial role of resonant driving in protecting quantum coherence and enabling persistent synchronization, even in the presence of a non-trivial non-Markovian environment.
The connection between resonance, coherence preservation, and synchronization provides a simple physical  mechanism that can be exploited for robust control of driven open quantum systems across different settings.

\section*{Acknowledgments}
F.S. acknowledges support from Magnus Ehrnroothin S\"a\"ati\"o.
B.V. acknowledges support from MUR and Next Generation EU via the PRIN 2022 Project “Quantum Reservoir Computing (QuReCo)” (contract n. 2022FEXLYB)  and the NQSTI-Spoke1-BaC project QSynKrono (contract n. PE00000023-QuSynKrono).

\bibliography{My_Library,bib_new}

\end{document}